%% file: HIN-14-004_temp.tex
\pdfoutput=1

\documentclass[11pt,twoside,a4paper,cmspaper,final,collab]{cms-tdr}
\def\svnVersion{344468}\def\cmsCernNoTag{CERN-PH-EP/2015-209}\def\cmsCernDate{\today}\def\cmsMessage{Published in Physical Review Letters as \href{http://dx.doi.org/10.1103/PhysRevLett.116.032301}{\doi{10.1103/PhysRevLett.116.032301.}}}

\begin{document}\cmsNoteHeader{HIN-14-004}

\hyphenation{had-ron-i-za-tion}
\hyphenation{cal-or-i-me-ter}
\hyphenation{de-vices}
\RCS$HeadURL: svn+ssh://svn.cern.ch/reps/tdr2/papers/HIN-14-004/trunk/HIN-14-004.tex $
\RCS$Id: HIN-14-004.tex 344360 2016-05-26 04:40:01Z hckim $
\newlength\cmsFigWidth
\ifthenelse{\boolean{cms@external}}{\setlength\cmsFigWidth{0.85\columnwidth}}{\setlength\cmsFigWidth{0.4\textwidth}}
\ifthenelse{\boolean{cms@external}}{\providecommand{\cmsLeft}{top}}{\providecommand{\cmsLeft}{left}}
\ifthenelse{\boolean{cms@external}}{\providecommand{\cmsRight}{bottom}}{\providecommand{\cmsRight}{right}}

\cmsNoteHeader{HIN-14-004}
\newcommand{\sqrts}{ \ensuremath{\sqrt{s}}\xspace}
\newcommand{\sqrtsNN}{ \ensuremath{\sqrt{s_{_{\mathrm{NN}}}}}\xspace}
\providecommand{\PBz}{\ensuremath{{\cmsSymbolFace{B}}{}^0}\xspace}
\providecommand{\PBzs}{\ensuremath{{\cmsSymbolFace{B}}{}^0_\cmsSymbolFace{s}}\xspace}
\providecommand{\PABz}{\ensuremath{\overline{\cmsSymbolFace{B}}{}^0}\xspace}
\providecommand{\PABzs}{\ensuremath{\overline{\cmsSymbolFace{B}}{}^0_\cmsSymbolFace{s}}\xspace}
\newcommand{\Bplusdecay}{\ensuremath{\PBp\rightarrow\PJGy~\PKp\rightarrow\Pgmp\Pgmm\PKp}\xspace}
\newcommand{\Bzerodecay}{\ensuremath{\PBz\rightarrow\PJGy~\PKst\rightarrow\Pgmp\Pgmm\PKp\Pgpm}\xspace}
\newcommand{\kstardecay}{\ensuremath{\PKst\rightarrow\PKp\Pgpm}\xspace}
\newcommand{\phidecay}{\ensuremath{\phi\rightarrow\PKp\PKm}\xspace}
\newcommand{\Bsubsdecay}{\ensuremath{\PBzs\rightarrow\PJGy~\phi \rightarrow\Pgmp\Pgmm\PKp\PKm}\xspace}
\newcommand{\Jpsidecay}{\ensuremath{\PJGy\rightarrow\Pgmp\Pgmm}\xspace}
\newcommand{\Bjpsixdecay}{\ensuremath{\PB\rightarrow\PJGy~\cmsSymbolFace{X}}\xspace}
\newcommand{\Bzerobgdecay}{\ensuremath{\PBz\rightarrow\PJGy~\PKp\Pgpm}\xspace}
\newcommand{\DecayKPi}{\ensuremath{\PKp\Pgpm}\xspace}
\newcommand{\DecayKK}{\ensuremath{\PKp\PKm}\xspace}
\newcommand{\RpPb}{\ensuremath{R_\mathrm{pA}}\xspace}
\newcommand{\ylab}{\ensuremath{y_\text{lab}}\xspace}
\newcommand{\ycm}{\ensuremath{y_\mathrm{CM}}\xspace}
\newcommand{\RpAfonllvspt}{\RpPb^\mathrm{FONLL}(\pt)}
\newcommand{\RpAfonll}{\RpPb^\mathrm{FONLL}}
\newcommand{\RpAfonllnopt}{\RpPb^\mathrm{FONLL}}
\newcommand{\RpAfonllvsy}{\RpPb^\mathrm{FONLL}(\ycm)}
\newcommand{\Pb}{\ensuremath{\mathrm{Pb}}\xspace}
\newcommand{\pPb}{\ensuremath{\mathrm{pPb}}\xspace}
\newcommand{\PbPb}{\ensuremath{\mathrm{PbPb}}\xspace}
\newcommand{\pp}{\ensuremath{\mathrm{pp}}\xspace}
\newcommand{\mcscale}{m_{c}}
\newcommand{\mbscale}{m_{b}}
\newcommand{\mufscale}{\mu_{R}}
\newcommand{\murscale}{\mu_{F}}
\newcommand{\muzeroscale}{\mu_{0}}
\newcommand{\masst}{m_{t}}
\newcommand{\Bzerodecaynofinal}{\ensuremath{\PBz\rightarrow\PJGy~\PKst}\xspace}

\title{\texorpdfstring{Study of $\PB$ meson production in $\pPb$ collisions at $\sqrtsNN=5.02\TeV$ using exclusive hadronic decays}{Study of B meson production in pPb collisions at sqrt(s[NN])=5.02 TeV using exclusive hadronic decays}}

\date{\today}

\abstract{
The production cross sections of the $\PBp$, $\PBz$, and $\PBzs$ mesons, and of their charge conjugates, are measured via exclusive hadronic decays in $\pPb$
collisions at the center-of-mass energy $\sqrtsNN=5.02\TeV$ with the CMS detector at the CERN LHC. The dataset used for this analysis corresponds to an integrated luminosity of $34.6\nbinv$.
The production cross sections are measured in the transverse momentum range between 10 and 60\GeVc. No significant modification is observed compared to proton-proton perturbative QCD calculations scaled by the number of incoherent nucleon-nucleon collisions. These results provide a baseline for the study of
in-medium $\cPqb$ quark energy loss in $\PbPb$ collisions.}

\hypersetup{
  pdfauthor={CMS Collaboration},
  pdftitle={Study of B meson production in pPb collisions at sqrt(s[NN])=5.02 TeV using exclusive hadronic decays},
  pdfsubject={CMS},
  pdfkeywords={physics, dimuons, proton Lead, charmonia, suppression, quark gluon plasma, shadowing, B~meson, open heavy-flavor}}

\maketitle

Relativistic heavy ion collisions allow the study of quantum chromodynamics (QCD) at very high temperature and density.
Under such extreme conditions, a strongly interacting state consisting of deconfined quarks and gluons, the quark-gluon plasma (QGP)~\cite{QGP1,QGP2}, is predicted by lattice QCD calculations~\cite{Karsch:2003jg}. Hard-scattered partons are expected to lose energy as they traverse the QGP via elastic collisions and medium-induced gluon radiation. The resulting reduction of the measured yield of hadrons, compared to expectations based on proton-proton ($\pp$) data, is often referred to as "jet quenching"~\cite{Eloss1,Eloss2}. The flavor dependence of jet quenching is one of the most important testing grounds for energy loss
models~\cite{Dokshitzer:2001zm,Armesto:2003jh,Wicks:2007am,Zhang:2003wk,Adil:2006ra}. However, other phenomena can affect the yield of heavy-flavor particles,
independently of the presence of a deconfined partonic medium. For instance, modifications of the parton distribution functions (PDFs) in the nucleus with respect to nucleon PDFs~\cite{Eskola:2009uj,deFlorian:2003qf,Frankfurt:2011cs} could change the production rate.
Therefore, a complete understanding of the interactions of heavy quarks in the deconfined medium formed in heavy ion collisions requires a thorough knowledge of their production in proton- (or deuteron-) nucleus,  p(d)A, collisions.

Currently, published data for heavy flavor production in p(d)A exist for open charm both at RHIC, in dAu collisions at $\sqrtsNN = 200\GeV$, and at the LHC in $\pPb$ collisions
at $\sqrtsNN = 5.02\TeV$. At RHIC, the STAR Collaboration measured the charm spectra in the rapidity interval $\abs{y}<1$ from direct reconstruction of $\PDz$ meson
and from indirect electron and positron measurements of charm semileptonic decays~\cite{STAR:prl94}. The measured yields were found to be consistent within the uncertainties with the hypothesis of binary scaling (no modification with respect to nucleon-nucleon (NN) cross section scaled by the number of incoherent NN binary collisions). However, the PHENIX Collaboration measured a significant enhancement of the production of heavy-flavor decay electrons in $\abs{y}<0.35$ in high-multiplicity dAu events with respect to a combined, data and theory, $\pp$ reference~\cite{PHENIX:prl109}. Recently, PHENIX also measured a significant enhancement of heavy-flavor production via single-muon detection
at backward rapidity (the Au-going direction), and a suppression at forward rapidities (the d-going direction)~\cite{PHENIX:ax1310}.
This measured difference in heavy-flavor production between forward and backward rapidities is significantly larger than predicted by leading-order
perturbative QCD calculations with nuclear PDFs~\cite{Helenius:2012uj}. In \pPb collisions at the LHC, the ALICE Collaboration measured the production of the $\PD$~meson in the $-0.96<y<0.04$ interval, and found it to be, within uncertainties, compatible with $\pp$ data scaled by the number of binary NN collisions, over a large transverse momentum (\pt) range~\cite{Abelev:2014hha}. The LHC results are well described by theoretical calculations that do not require a deconfined medium to be formed in the collision. This supports the idea that the $\PD$~meson suppression at high-\pt observed in $\PbPb$ collisions by the ALICE Collaboration~\cite{Adam:2015nna} is due to parton interactions with the deconfined medium. While measurements, both at RHIC and LHC, support that most of the suppression observed in AA collisions is due to partonic energy loss, the details of the phenomena affecting open charm in pA and AA collisions are still to be understood.

The production of $\PB$~mesons was studied at the LHC in proton-proton ($\pp$) collisions at $\sqrts = 7$\TeV over wide \pt and rapidity intervals by CMS~\cite{CMSBmesonpp,Chatrchyan:2011pw,Chatrchyan:2011vh}, ATLAS~\cite{ATLAS:2013cia}, and LHCb~\cite{LHCb:2013JHEP}. In $\PbPb$ collisions, CMS measured the non-prompt $\PJGy$ from $\PB$ hadron decays at $\sqrtsNN = 2.76\TeV$~\cite{CMSNonPromptJpsi}, and observed a strong suppression with respect to the hypothesis of binary scaling.
In this Letter, we extend the study of heavy-quark production in p(d)A collisions by performing the first measurement of exclusive $\PB$~meson decays in $\pPb$ collisions.

The $\PB$~mesons are measured in a region $\abs{y_\text{lab}}<2.4$ via the full reconstruction of their decay
channels: $\Bplusdecay$ with branching fraction  $\mathcal{B} = (6.12 \pm 0.19) \times 10^{-5}$, $\Bzerodecay$ with $\mathcal{B}=(5.24 \pm 0.24) \times 10^{-5}$, and $\Bsubsdecay$ with $\mathcal{B} = (3.12 \pm 0.27) \times 10^{-5}$~\cite{PDG}. As this analysis does not separate $\PBp$ from $\PBm$, $\PBz$ from $\PABz$, or $\PBzs$ from $\PABzs$, mesons are referred to generically as $\PBp$, $\PBz$, and $\PBzs$, respectively, for the purposes of reconstruction. For the final cross section values, the combined results are divided by two to obtain an average.

The CMS detector has excellent capabilities to reconstruct $\PB$~meson decays due to the highly efficient muon detection system and the high-resolution
silicon tracker~\cite{bib_CMS}. The data sample used in this analysis corresponds to an integrated luminosity of $(34.6 \pm 1.2)\nbinv$~\cite{LumiPAS}.
The direction of the proton beam was initially opposite to the positive direction of the CMS longitudinal axis~\cite{bib_CMS},
and it was reversed after 60\% of the data were taken. The beam energies were $4\TeV$ for protons and $1.58\TeV$ per nucleon for lead nuclei, resulting in a nucleon-nucleon center-of-mass energy of $\sqrtsNN = 5.02\TeV$. Because of the energy difference of the colliding beams,
the nucleon-nucleon center-of-mass frame in $\pPb$ collisions was not at rest with respect to the laboratory frame. The results presented here use the convention that the proton-going side corresponds to positive pseudorapidity. This implies that massless particles emitted at pseudorapidity $\eta_\mathrm{CM}$ in the NN center-of-mass frame are detected at $\eta_\text{lab} = \eta_\mathrm{CM} + 0.465$.

A detailed description of the CMS experiment and coordinate system can be found in Ref.~\cite{bib_CMS}. Only the detector subsystems most relevant for this analysis are described here. Charged particles (tracks) are reconstructed within the range $\abs{\eta_\text{lab}}< 2.5$  by using
the silicon tracker detector, located in the 3.8\unit{T} magnetic field of a superconducting solenoid.
Muons are identified in the interval $\abs{\eta_\text{lab}}<2.4$ with gas-ionization detectors made of three technologies:
drift tubes, cathode strip chambers, and resistive plate chambers, embedded in the steel flux-return yoke of the magnet. The CMS apparatus also has extensive forward calorimetry, including two steel and quartz-fiber Cherenkov hadron forward (HF) calorimeters, which cover the range $2.9 <\abs{\eta_\text{lab}}<5.2$.

Events used in the measurement are collected with a trigger requiring the presence of a muon with $\pt >3\GeVc$.
To select inelastic hadronic interactions, the offline analysis requires a coincidence of at least one of the HF calorimeter towers (with more than 3\GeV of total energy) from each side of the interaction point. Events are further required to have at least one reconstructed primary vertex, formed by at least two tracks, with a distance from the center of the nominal interaction region of less than $15\unit{cm}$ along the beam axis.

Several Monte Carlo (MC) simulated event samples are used to evaluate background components and signal efficiencies, specifically: i) an inclusive (prompt and non-prompt) $\PJGy$ sample; ii) a sample containing all $\PB$~mesons decaying into a $\PJGy$; iii) a signal-only sample with the $\PBp$, $\PBz$, and $\PBzs$ decays included in the present analysis. First, proton-proton collisions are simulated with \PYTHIA 6.424~\cite{pythia} tune Z2~\cite{Field:2010bc} and propagated through the CMS detector using the \GEANTfour package~\cite{geant4}. The $\PB$~meson decays are simulated with the \EVTGEN package~\cite{evtgen}, and final state photon radiation in the $\PB$ decays  is simulated by \PHOTOS~\cite{Barberio:1990ms}. Then, the \PYTHIA events are embedded into simulated $\pPb$ events produced by the {\sc hijing} generator version 1.383~\cite{hijing}, which is tuned to reproduce global event properties such as charged-hadron \pt spectra and particle multiplicity.

Muons are required to be within the following kinematic region: $\pt^{\mu}>3.3\GeVc$ for $\abs{\eta^{\mu}_\text{lab}}<1.3$, total momentum $p^{\mu} > 2.9\GeVc$ for $1.3 < \abs{\eta^{\mu}_\text{lab}} < 2.2$, or $\pt^{\mu} > 1.5\GeVc$ for $2.2 < \abs{\eta^{\mu}_\text{lab}} < 2.4$~\cite{Khachatryan:2010yr}. This acceptance selection is chosen so as to guarantee a single-muon detection probability exceeding about 10\%. Two muons of opposite charge with an invariant mass within $150\MeVcc$ of the world-average $\PJGy$ mass~\cite{PDG} are selected to reconstruct a $\PJGy$ candidate, with a mass resolution of typically 18--55\MeVcc, degrading as a function of the dimuon rapidity. The $\PB$~meson candidates are formed by combining $\PJGy$ candidates with
charged tracks. Without using particle identification, assumptions need to be made about the masses of the charged tracks. In calculating the mass of the $\PBp$ candidates, the single charged particle is always assumed to have the mass of a kaon. In the $\PBz$ case, two invariant mass values are computed, corresponding to the two possible assignments of the kaon and pion masses to the two-track system. For $\PBzs$ candidates, the two charged tracks are always assumed to be kaons.
Single track low $\pt$ thresholds of 0.9, 0.7, and 0.4\GeVc are applied in the $\PBp$, $\PBz$, and $\PBzs$ analyses,
respectively, to reduce the combinatorial background, which is further minimized by additional selection criteria.
In particular, $\PB$ candidates are selected according to the $\chi^{2}$ probability of the decay vertex (the probability for the $\PJGy$ muon tracks and the other charged track to point to a common vertex), the 3D flight distance (normalized by its uncertainty) between the primary and decay vertices, and the pointing angle, which is defined as the angle between the line connecting the primary and decay vertices and the momentum vector of the $\PB$~meson in the plane transverse to the beam direction.
The selection is optimized for each meson species using a multivariate technique that uses the genetics algorithm~\cite{TMVA},
in order to maximize the statistical significance of the $\PB$~meson signals. In the $\PBz$ and $\PBzs$ analyses, the invariant masses of the $\DecayKPi$ and the $\DecayKK$ are required to be compatible with the masses of the $\cmsSymbolFace{K^{0\ast}(892)}$, $\PKst$ and the $\phi$ resonances, respectively. If more than one candidate in a given event survives all the aforementioned selection criteria, the candidate with the best vertex $\chi^{2}$ probability is selected.

\begin{figure*}[tb]
\centering
\includegraphics[width=.99\textwidth]{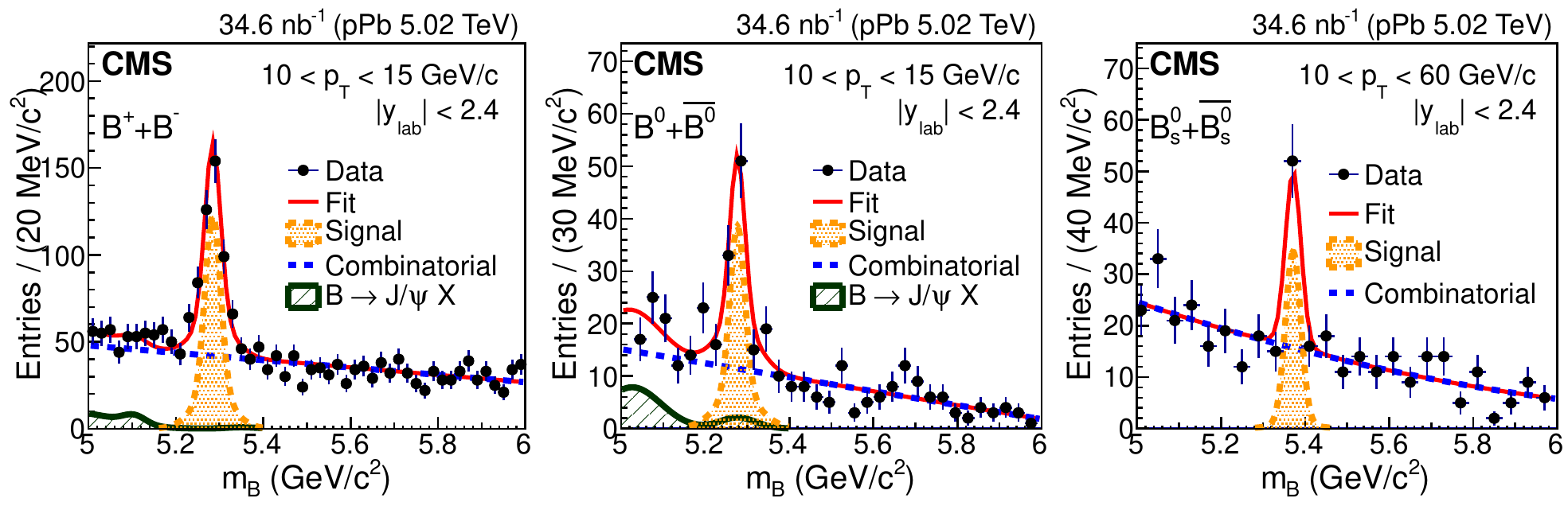}
\caption{Invariant mass distributions of $\PBp+\PBm$ (left), $\PBz+\PABz$ (center),  and $\PBzs+\PABzs$ (right) candidates in
the transverse momentum regions 10--15, 10--15, and 10--60\GeVc, respectively. See the text for details.}
\label{fig:rawYieldsBmeson}

\end{figure*}

The raw yields of $\PBp$, $\PBz$, and $\PBzs$ are extracted using a binned maximum likelihood fit to
the $\PB$~meson invariant-mass distributions in the mass range $5<\rm{m}_{\PB} <6\GeVcc$.
The invariant mass distributions of $\PBp$, $\PBz$, and $\PBzs$ candidates in the \pt regions 10--15, 10--15, and 10--60\GeVc, respectively, are shown in Fig.~\ref{fig:rawYieldsBmeson}. In the case of $\PBp$ and $\PBz$, this choice corresponds to the lowest \pt interval
used in the analysis, while for $\PBzs$ it is the only interval. The signal shape is modeled by two Gaussians with the same mean values
(a free parameter in the fit) and different widths determined in MC simulations. The background is dominated by random combinations of prompt and non-prompt $\PJGy$ candidates with extra particles. This combinatorial background is modeled by a first-order polynomial in the $\PBp$ and $\PBz$ analyses, and by a second-order polynomial in the $\PBzs$ analysis, as suggested by studies on the embedded inclusive $\PJGy$ sample.
The background component shown as a crosshatched histogram and labeled as $\Bjpsixdecay$ in Fig.~\ref{fig:rawYieldsBmeson} is due to
mis-reconstructed $\PB$~meson decays that produce broad peaking structures in the invariant
mass region below 5.4\GeVcc. As an example, in the $\PBp$ analysis, a peaking background structure is created by $\Bzerodecaynofinal$ decays in which one decay product is lost in the $\PB$ candidate reconstruction. These background sources are studied with the embedded MC sample including all $\PB$~meson decays into final states with a $\PJGy$, and found to be well described by a superposition of 4 and 2 Gaussian functions in the $\PBp$ and $\PBz$ analyses, respectively. The resulting functional form, with the overall normalization left floating, is included in the global fit function.
This background component is found to be negligible in the $\PBzs$ analysis as a consequence of the selection on the mass of
the $\phi$ candidate.

The $\pt$-differential production cross section of the various $\PB$~meson species is computed in each \pt interval:
\begin{linenomath}
\begin{equation}
  \left.\frac{{\rd}\sigma}{{\rd}\pt}\right|_{\abs{\ylab}<2.4} =
  \frac{1}{2}\frac{1}{\Delta\pt}\frac{N(\pt)_{\abs{\ylab}<2.4}}{(\text{Acc}\,\epsilon)_{\abs{\ylab}<2.4} \, \PB \, \mathcal{L}}.
  \label{eq:crosssectionDpPb}
\end{equation}
\end{linenomath}
$N(\pt)_{\abs{\ylab}<2.4}$ is the raw signal yield extracted in each \pt interval of width $\Delta \pt$, $(\text{ Acc}\,\epsilon)_{\abs{\ylab}<2.4}$
represents the corresponding acceptance times efficiency, $\mathcal{L}$ is the integrated luminosity,
and $\PB$ is the branching fraction of the decay chain.
The factor 1/2 accounts for the fact that the yields were measured for particles and antiparticles added together,
but the cross section is given for particles only. An analogous expression holds for the
rapidity-differential cross section. The $(\text{Acc}\,\epsilon)$ correction factors are evaluated using
\textsc{pythia+hijing} simulations in each $\pt$ and $\abs{\ylab}$ interval, to account for the loss of signal due to the detector coverage, and to the trigger, reconstruction and offline selection. They vary, over the measured \pt and $y$ intervals, from 9\%to 37\% (3\% to 15\%) for $\PBp$ ($\PBz$), and they equal 8\% in the single bin used for the $\PBzs$.

\begin{figure*}[tb]
\centering
\includegraphics[width=.99\textwidth]{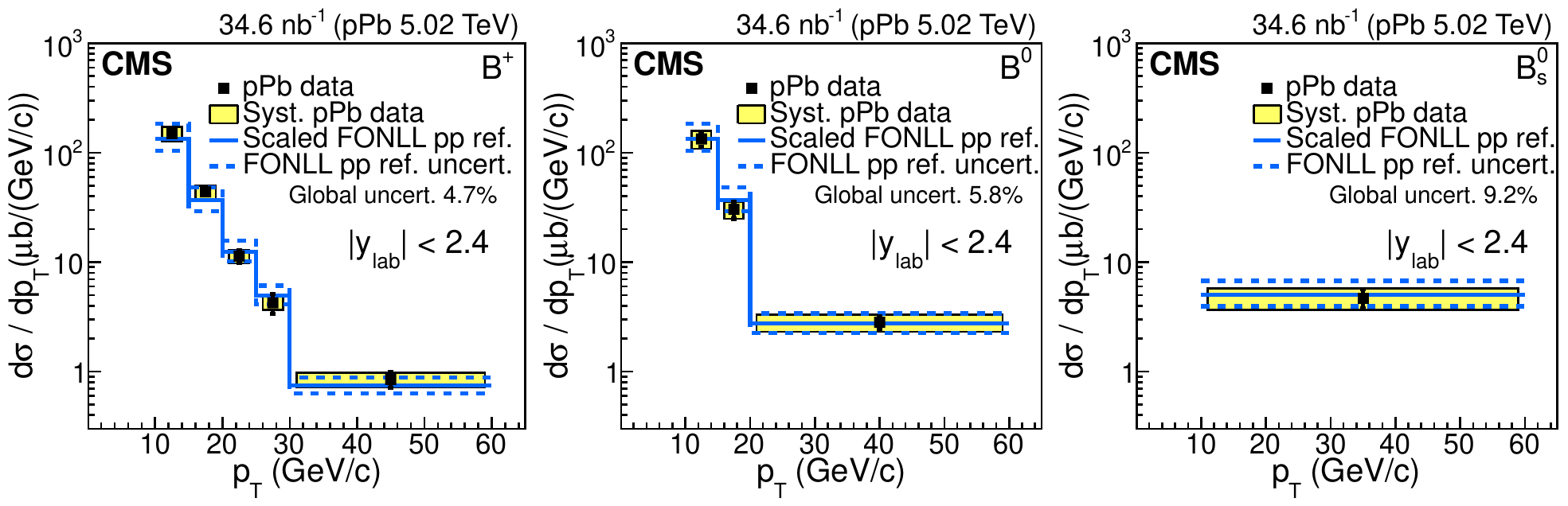}
\caption{The $\pt$-differential production cross section of $\PBp$ (left), $\PBz$ (center), and $\PBzs$ (right) measured in $\pPb$ collisions
at $\sqrtsNN=5.02\TeV$. The vertical bars (boxes) correspond to statistical (systematic) uncertainties.
The global systematic uncertainty, listed in each panel and not included in the data points, comprises the uncertainties in the integrated luminosity measurement and the $\PB$~meson
branching fractions. Results are compared to FONLL calculations~\cite{FONLLcharmbottomPP1,FONLLcharmbottomPP2,FONLLcharmbottomPP3}, scaled by the number of binary NN collisions, represented by a continuous histogram.
The dashed histograms represent the theoretical uncertainties for the FONLL reference.}
\label{fig:crosssections}

\end{figure*}

The cross sections are affected by several sources of systematic uncertainties arising from the signal extraction,
acceptance and efficiency corrections, branching fractions, and integrated luminosity determination.
The uncertainty from the fitting procedure (varying from 10\% to 15\% across all analysis intervals, for all three mesons) is evaluated
by varying the probability distribution functions used to model the signal and background distributions. As an alternative combinatorial background shape,
a second-order polynomial is used for $\PBp$ and $\PBz$, and a third-order polynomial for $\PBzs$. The uncertainty on the signal is evaluated
by considering three fit variations: (i) leaving free the width parameters; (ii) varying the width parameters by $\pm$20\% with respect to the MC value; (iii) using only one Gaussian. The maximum of all variations is propagated as systematic uncertainty, and it is given in the case of $\PBp$
by variation (i), and by variation (ii) for the other two mesons. The systematic uncertainties associated with the bin-by-bin acceptance correction (0.2\% to 5.6\%) are estimated by varying the shape of the generated $\PB$~meson \pt and $y$ spectra within limits defined by differences (including their statistical uncertainties) between data and MC calculations. For all three mesons, the $\PBp$ and $\PBz$ \pt spectrum shapes are assumed, while only the $\PBp$ is used for the $y$ shape. Using these shape variations, simplified (``toy") MC simulations are used to recalculate the acceptance in each kinematic bin, the maximum variation between the nominal acceptance and the toys being propagated as the systematic uncertainty. The systematic uncertainty due to the selection of the $\PB$~meson candidates (4\% to 11\%) is equal to 1 minus the ratio of the selection efficiencies (the ratio of the extracted yield with and without applying the selection) estimated in data and simulation. In addition, an uncertainty associated with the accuracy of the best candidate selection (3\%) , which depends on the number of reconstructed $\PB$~meson candidates, is assigned. This is evaluated by reweighting the population of the {\sc pythia+hijing} events so that the distribution of the number of $\PB$~meson candidates per event matches the one from data. The uncertainties in the muon trigger, and muon track reconstruction and identification efficiencies (4.5\% to 7.3\%), are evaluated by using the ``tag-and-probe" technique~\cite{Khachatryan:2010xn} on \pPb data and the embedded MC sample. The systematic uncertainty associated with the track reconstruction efficiency (3.9\% per hadronic track ~\cite{Chatrchyan:2014fea}) is estimated from a comparison of two-body and four-body $\PDz$ decays in pp data and MC calculations, all samples being reconstructed with the same tracking algorithm as the pPb sample.
The systematic uncertainty in the cross section measurement is computed point-by-point as the sum in quadrature of the different contributions mentioned above. In addition, a global systematic uncertainty is calculated to account for the uncertainties in the integrated luminosity value (3.5\%~\cite{LumiPAS}),
and in the $\PB$~meson branching fractions (3.1\%, 4.6\%, and 8.7\% for $\PBp$, $\PBz$, and $\PBzs$, respectively~\cite{PDG}).

\begin{figure*}[tb]
\centering
\includegraphics[width=.99\textwidth]{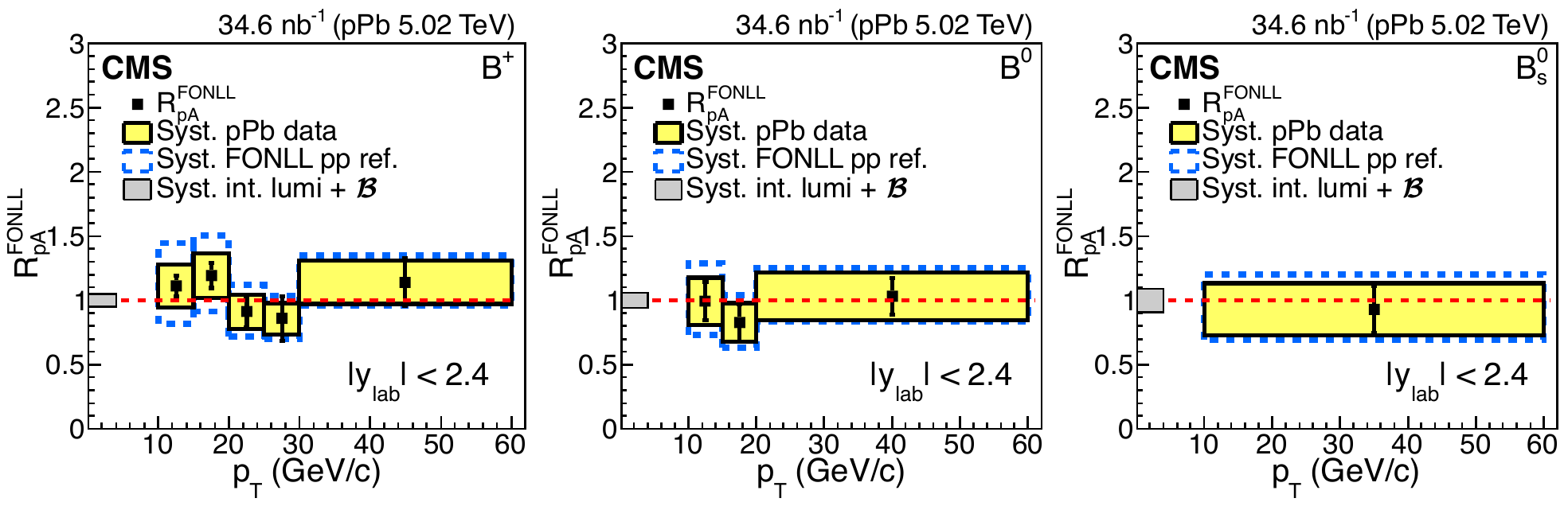}
\caption{The nuclear modification factors $\RpAfonllvspt$ of $\PBp$ (left), $\PBz$ (center), $\PBzs$ (right) measured in $\pPb$
collisions at $\sqrtsNN=5.02\TeV$. The statistical and systematic uncertainties on the $\pPb$ data are shown as bars and yellow boxes around the data points, respectively.
The systematic uncertainties from the FONLL predictions are plotted separately as open blue boxes.
The global systematic uncertainties are shown as full grey boxes at unity, and are not included in the data points.}
\label{fig:rpaall}

\end{figure*}

In Fig.~\ref{fig:crosssections}, the $\pt$-differential production cross sections of all three $\PB$~mesons measured in the interval $\abs{\ylab} < 2.4$ are presented, with data points placed at the center of each bin. They are compared to the $\pp$ cross sections obtained from fixed-order plus next-to-leading-logarithm (FONLL) calculations~\cite{FONLLcharmbottomPP3}, which reproduce the $\PB$~meson \pt-differential cross sections in $\pp$ collisions at 7 TeV~\cite{CMSBmesonpp,Chatrchyan:2011pw,Chatrchyan:2011vh,ATLAS:2013cia,LHCb:2013JHEP}. The individual cross sections are obtained by scaling the FONLL total beauty production~\cite{FONLLcharmbottomPP1,FONLLcharmbottomPP2,FONLLcharmbottomPP3} by the world-average production fractions of $\PBp$, $\PBz$, and $\PBzs$ (40.2\%, 40.2\%, and 10.5\%, respectively~\cite{PDG}). The obtained $\PBp$ FONLL reference is validated using published experimental cross sections measured in $\pp$ collisions at $\sqrt{s}= 7$\TeV~\cite{CMSBmesonpp,ATLAS:2013cia}. The FONLL predictions are scaled by A($=208$), the atomic mass of the $\Pb$ nucleus, to account for the number of binary NN collisions~\cite{Miller:2007ri}. The FONLL uncertainties, which are larger than the experimental uncertainties, represent the quadratic sum of several variations made to the calculation: of the factorization and renormalization scales, of the $\cPqb$ quark mass, and of the uncertainty associated with PDFs (providing the largest contribution)~\cite{FONLLcharmbottomPP1,FONLLcharmbottomPP2,FONLLcharmbottomPP3}.
The nuclear modification factor $\RpAfonll$, shown in Fig.~\ref{fig:rpaall}, is computed as:
\begin{linenomath}
\begin{equation}
 \label{eq:RpAdefinition}
{R^\mathrm{FONLL}_{\rm pA}}=\frac{({\frac{{\rd}\sigma}{{\rd}\pt}})_{\rm pPb}}{\mathrm{A} \, ({\frac{{\rd}\sigma}{{\rd}\pt}})^\mathrm{FONLL}_{\pp}}
\end{equation}
\end{linenomath}
where the numerator is defined in Eq.~\ref{eq:crosssectionDpPb} and the denominator is the corresponding theoretical calculation for $\PB$~meson production in $\pp$ collisions at the same center-of-mass energy. The theoretical uncertainties represented by the open blue boxes in Fig.~\ref{fig:rpaall} are computed by recalculating $\RpAfonllvspt$ with the upper
and lower values of the FONLL predictions \hyphenation{re-fe-rence} represented by dashed histograms in Fig.~\ref{fig:crosssections}.
The nuclear modification factors of the three $\PB$~mesons do not show evidence for modification of $\pPb$ data compared to the FONLL reference, in the considered \pt range within the quoted uncertainties. No significant differences are observed between the three $\PB$~meson species. In the lowest \pt interval measured, $\RpAfonllvspt$ is $1.11 \pm 0.08\stat\pm 0.17\,\text{(syst \pPb)}\,_{- 0.29}^{+0.33}\,\text{(syst  FONLL)}$,  $0.99 \pm 0.15\stat\pm 0.18\, \text{(syst  \pPb)}\, _{- 0.26}^{+ 0.30}\,\text{(syst  FONLL)}$, and $0.93 \pm 0.18\stat\pm 0.20\, \text{(syst  \pPb)}\,_{- 0.24}^{+0.27}\,\text{(syst  FONLL)}$, for \PBp, \PBz, and \PBzs, respectively.

\begin{figure*}[!htb]
\centering
\includegraphics[width=.49\textwidth]{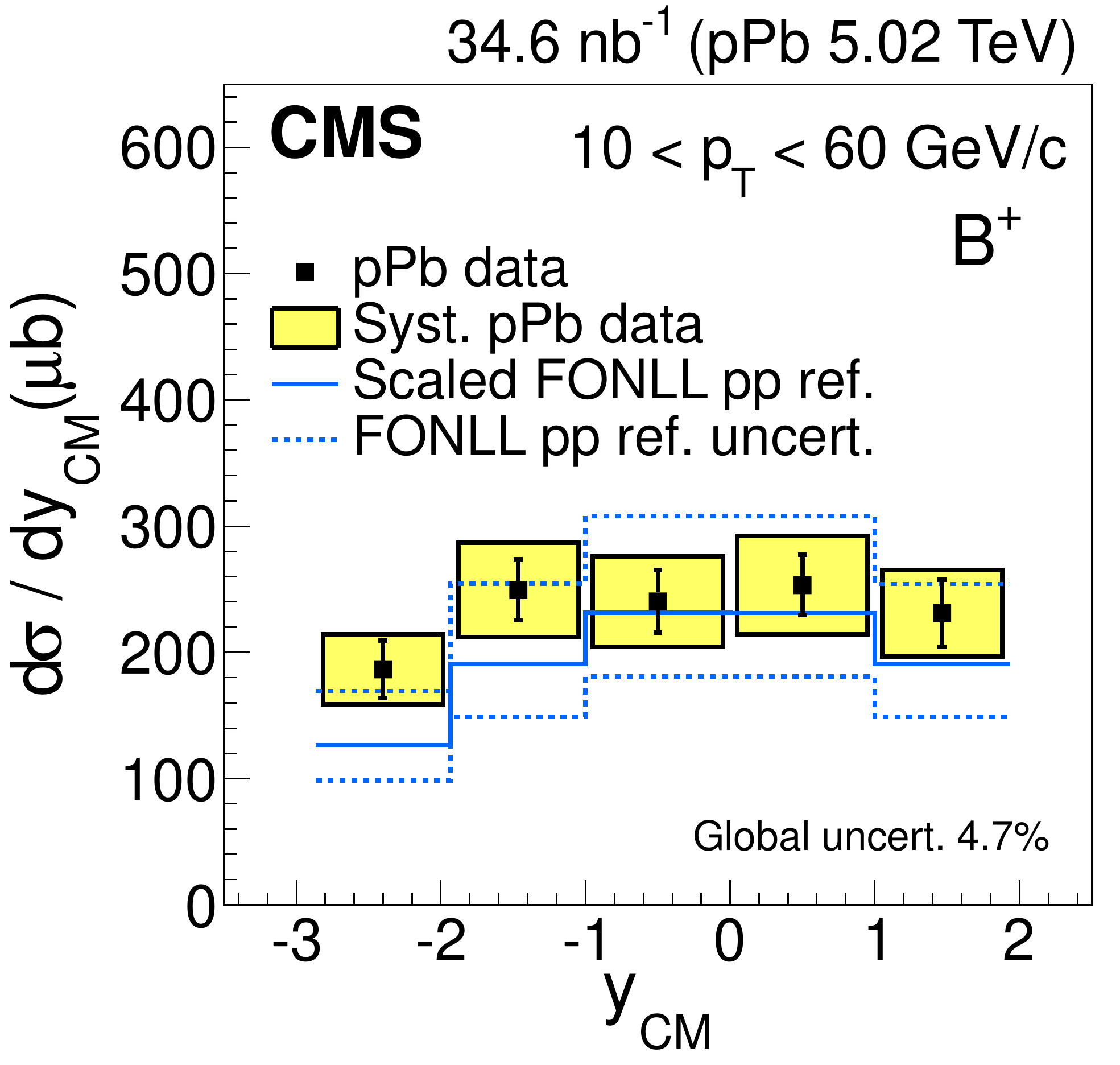}
\includegraphics[width=.49\textwidth]{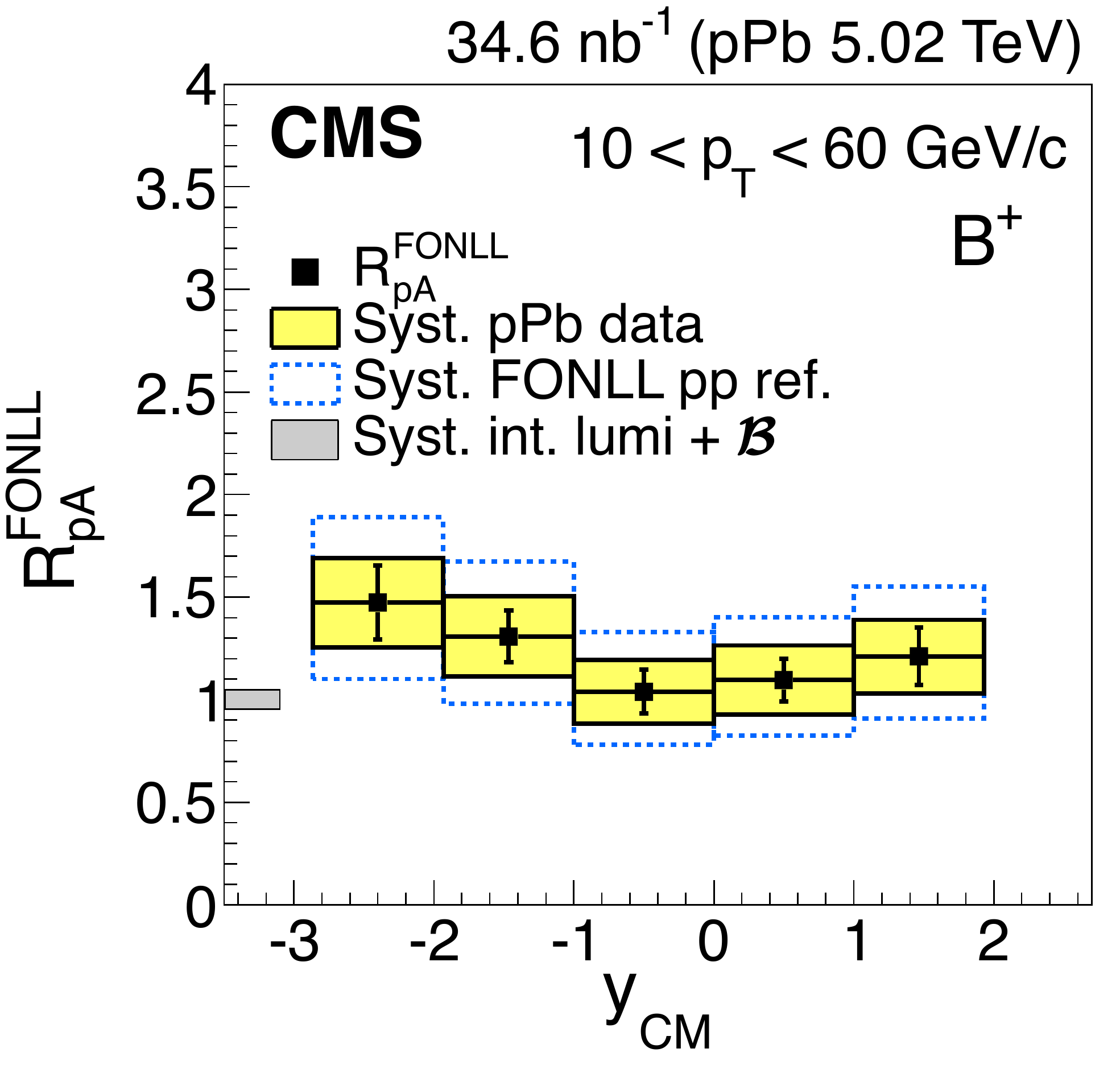}
\caption{(left) The $\ycm$-differential production cross section of $\PBp$ measured in $\pPb$ collisions at
$\sqrtsNN=5.02$\TeV. Vertical bars (the boxes) correspond to statistical (systematic) uncertainties.
The listed global systematic uncertainty is not included in the data points.
The result is compared to a FONLL calculation~\cite{FONLLcharmbottomPP1,FONLLcharmbottomPP2,FONLLcharmbottomPP3} represented by a continuous histogram. The dashed histograms represent the theoretical uncertainties for the FONLL reference. (right) The nuclear modification factor $\RpAfonllvsy$ of $\PBp$ as a function of \ycm.
The statistical and systematic uncertainties on the $\pPb$ data are shown as bars and yellow boxes around the data points, respectively. The systematic uncertainty from the FONLL reference is plotted separately as open blue boxes. The global systematic uncertainty is shown as a full grey box at unity, and is not included in the data points.}
\label{fig:crosssectionsY}

\end{figure*}

The production cross section of $\PBp$ is also studied as a function of its rapidity in the
center-of-mass frame (\ycm). The $\ycm$-differential cross section of $\PBp$ in the interval $10< \pt <60\GeVc$ is shown in Fig.~\ref{fig:crosssectionsY} (left). In Fig.~\ref{fig:crosssectionsY} (right), the rapidity dependence of the nuclear modification factor of $\PBp$ is shown. No strong evidence of rapidity dependence of $\RpAfonll$ is observed within the uncertainties.

In summary, the first measurements of the $\PBp$, $\PBz$, and $\PBzs$ meson production cross sections in \pPb\
collisions at $\sqrtsNN=5.02$\TeV are presented. The mesons are measured in $\abs{y_\text{lab}}<2.4$ and
$10<\pt<60\GeVc$ via the reconstruction of one of their exclusive hadronic decay channels. Within the transverse momentum and rapidity ranges studied,
no significant modifications are observed, considering the statistical and systematical uncertainties, when compared to $\pp$ FONLL calculations scaled by the number of incoherent nucleon-nucleon collisions.
These results provide a baseline for the study of in-medium $\cPqb$ quark energy loss in $\PbPb$ collisions.

\begin{acknowledgments}
We congratulate our colleagues in the CERN accelerator departments for the excellent performance of the LHC and thank the technical and administrative staffs at CERN and at other CMS institutes for their contributions to the success of the CMS effort. In addition, we gratefully acknowledge the computing centers and personnel of the Worldwide LHC Computing Grid for delivering so effectively the computing infrastructure essential to our analyses. Finally, we acknowledge the enduring support for the construction and operation of the LHC and the CMS detector provided by the following funding agencies: BMWFW and FWF (Austria); FNRS and FWO (Belgium); CNPq, CAPES, FAPERJ, and FAPESP (Brazil); MES (Bulgaria); CERN; CAS, MoST, and NSFC (China); COLCIENCIAS (Colombia); MSES and CSF (Croatia); RPF (Cyprus); MoER, ERC IUT and ERDF (Estonia); Academy of Finland, MEC, and HIP (Finland); CEA and CNRS/IN2P3 (France); BMBF, DFG, and HGF (Germany); GSRT (Greece); OTKA and NIH (Hungary); DAE and DST (India); IPM (Iran); SFI (Ireland); INFN (Italy); MSIP and NRF (Republic of Korea); LAS (Lithuania); MOE and UM (Malaysia); CINVESTAV, CONACYT, SEP, and UASLP-FAI (Mexico); MBIE (New Zealand); PAEC (Pakistan); MSHE and NSC (Poland); FCT (Portugal); JINR (Dubna); MON, RosAtom, RAS and RFBR (Russia); MESTD (Serbia); SEIDI and CPAN (Spain); Swiss Funding Agencies (Switzerland); MST (Taipei); ThEPCenter, IPST, STAR and NSTDA (Thailand); TUBITAK and TAEK (Turkey); NASU and SFFR (Ukraine); STFC (United Kingdom); DOE and NSF (USA).
\end{acknowledgments}
\bibliography{auto_generated}
\cleardoublepage \appendix\section{The CMS Collaboration \label{app:collab}}\begin{sloppypar}\hyphenpenalty=5000\widowpenalty=500\clubpenalty=5000\input{HIN-14-004-authorlist.tex}\end{sloppypar}
\end{document}

%% file: HIN-14-004-authorlist.tex
\textbf{Yerevan Physics Institute,  Yerevan,  Armenia}\\*[0pt]
V.~Khachatryan, A.M.~Sirunyan, A.~Tumasyan
\vskip\cmsinstskip
\textbf{Institut f\"{u}r Hochenergiephysik der OeAW,  Wien,  Austria}\\*[0pt]
W.~Adam, E.~Asilar, T.~Bergauer, J.~Brandstetter, E.~Brondolin, M.~Dragicevic, J.~Er\"{o}, M.~Flechl, M.~Friedl, R.~Fr\"{u}hwirth\cmsAuthorMark{1}, V.M.~Ghete, C.~Hartl, N.~H\"{o}rmann, J.~Hrubec, M.~Jeitler\cmsAuthorMark{1}, V.~Kn\"{u}nz, A.~K\"{o}nig, M.~Krammer\cmsAuthorMark{1}, I.~Kr\"{a}tschmer, D.~Liko, T.~Matsushita, I.~Mikulec, D.~Rabady\cmsAuthorMark{2}, B.~Rahbaran, H.~Rohringer, J.~Schieck\cmsAuthorMark{1}, R.~Sch\"{o}fbeck, J.~Strauss, W.~Treberer-Treberspurg, W.~Waltenberger, C.-E.~Wulz\cmsAuthorMark{1}
% ...
\vskip\cmsinstskip
% ...
1:~~Also at Vienna University of Technology, Vienna, Austria\\
2:~~Also at CERN, European Organization for Nuclear Research, Geneva, Switzerland\\
% ...
